# Strain engineering of ion migration in LiCoO$_2$

Jia-Jing Li[1], Yang Dai[2*], Jin-Cheng Zheng[1,3,4*]

[1] Department of Physics, and Collaborative Innovation Center for Optoelectronic Semiconductors and Efficient Devices, Xiamen University, Xiamen 361005, China

[2] Department of Chemical Engineering, School of Environmental and Chemical Engineering, and Institute for Sustainable Energy, Shanghai University, Shangda Road 99, Shanghai, 200444, China

[3] Department of Physics, Xiamen University Malaysia, 439000 Sepang, Selangor, Malaysia

[4] Fujian Provincial Key Laboratory of Theoretical and Computational Chemistry, Xiamen University, Xiamen 361005, China

* Corresponding authors: dy1982@shu.edu.cn (DY), jczheng@xmu.edu.cn (JCZ)

**Abstract**

Strain engineering is a powerful approach for tuning various properties of functional materials. The influences of lattice strain on the Li-ion migration energy barrier of lithium-ions in layered LiCoO$_2$ have been systemically studied using lattice dynamics simulations, analytical function and neural network method. We have identified two Li-ion migration paths, oxygen dumbbell hop (ODH), and tetrahedral site hop (TSH) with different concentrations of local defects. We found that Li-ion migration energy barriers increased with the increase of pressure for both ODH and TSH cases, while decreased significantly with applied tensile uniaxial *c*-axis strain for ODH and TSH cases or compressive in-plane strain for TSH case. Our work provides the complete strain-map for enhancing the diffusivity of Li-ion in LiCoO$_2$, and therefore, indicates a new way to achieve better rate performance through strain engineering.

## I. INTRODUCTION

Lithium-ion batteries have been playing a paramount role in energy storage fields for more than three decades [1-10]. Challenges for rechargeable lithium batteries have been discussed [2,3], and progresses such as investigation of thermal instability [4], and utilization of nanomaterials [8] including carbon nanotubes [9], nanowires [11], and borophene [12] have been reported. The application of rechargeable batteries in portable electronic devices has been reviewed recently [13]. LiCoO$_2$, the early developed cathode material, is in possession of excellent structure stability, charge/discharge reversibility, outstanding electrical and ionic conductivities [1,6]. Furthermore, since LiCoO$_2$ has a perfect layered crystal structure with a close-packed oxygen-anion framework, the battery employing the LiCoO$_2$ cathode material has the highest volumetric energy density, finding the dominant application in electronic terminal devices [1,6,14]. Although the LiCoO$_2$ cathode material is already well developed, there still has rooms to improve the rate capability, safety and longevity.

Several material treatments such as doping [15-18], or surface coating [14] have been used to modify electronic properties or to improve stability. Applying strain is another effective way to tune the ionic conductivity of electrode materials. The diffusion path and energy barrier of Li ion in the solid materials are determined by the potential energy surface [7]. The Li ion migration potential energy surface can be tuned by the applied strain on the materials. In $LiCoO_2$, the diffusion and the intercalation/extraction of Li are mediated by a vacancy mechanism. The defect thermodynamics, mechanical properties and migration barriers for Li ions in $LiCoO_2$ are highly sensitive to mechanical strains [19-21].

Furthermore, the batteries always apply micrometer-sized single crystal $LiCoO_2$ to improve the volumetric energy density. The non-uniformity of the large particle during Li ion migration may induce considerable strain gradient. In addition, the aforementioned strategy of doping elements [15,18] in $LiCoO_2$ may also induce local lattice strain, due to the different ionic size of dopants. Especially, for the application of solid-state or thin film battery, the affections of strain will further amplify. Therefore, the strain has great effects on the migration of Li ions in $LiCoO_2$. Unfortunately, the coupling of strain field to the Li ion migration is less studied.

In our previous work, we have demonstrated that strain engineering is a powerful approach for tuning various properties of functional materials, such as ferromagnetism [22], superconductivity [23], thermal conductivity [24], and electron structure of 2D materials [25,26]. Therefore, we expect that the strain engineering method may be an important approach for tuning ion migration of $LiCoO_2$. In this work, we systematically studied the effects of lattice strain on the Li-ion migration energy barrier of $LiCoO_2$. Two possible Li ion migration mechanisms are explored. The strategy of strain engineering for reducing Li ion migration energy barrier has been given. Finally, additional data analysis such as empirical formula fitting and neutral network model has been proposed to predict the Li ion migration energy barrier under strain.

## II. METHODS

The lattice dynamics (LD) calculations have been performed using General Utility Lattice Program (GULP) [27]. The potential energies between ions were calculated using Born-Mayer-Huggins (BMH) potential, combining the long-range Coulombic component and short-range pair-wise interactions. The short-range interactions, $\Phi_{ij}$, were modeled using Buckingham potential:

$$\Phi_{ij}(r_{ij}) = A_{ij} \exp\left(-\frac{r_{ij}}{\rho_{ij}}\right) - \frac{C_{ij}}{r_{ij}^6} \qquad (1)$$

where $r_{ij}$ is the distance between ions *i* and *j*; *A*, *ρ* and *C* are empirical parameters. The so-called shell model [28,29] is also used to take account the effects of electronic polarization of transition metal and oxide ions. The parameters of Buckingham potential and core-shell interaction for LiCoO$_2$ were taken from Fisher et al [30], which have been shown to be able to reproduce experimental structural parameters accurately. Those Born-Mayer-Huggins (BMH) potentials have been used for computation of defect energies, surface energies, and Li-ion diffusion of LiCoO$_2$ cathode materials [30].

A 2x2x1 supercell has been used to create the initial position of defects, then the defect energy and ion migration were calculated by the Mott-Littleton method [31] as implemented in the GULP code. In this method, two spherical regions (region 1 and region 2a) were defined around the defect center, for inner and outer spherical shell of ions, respectively. Atoms outside of these spheres (region 2b) extend to infinity. The ions in regions 1 and 2 are assumed to be strongly and weakly perturbed by the defect, respectively. The ions in the inner spherical region are therefore relaxed explicitly. We found that the radii of 12 Å for region 1 (> 880 ions) and 15 Å for region 2 were large enough to achieve the good convergence of the defect energy for LiCoO$_2$.

Machine learning methods [32-37] have been developed with long history, and are often used to discover hidden, attractive, and potentially useful patterns and relationships from massive data [36]. For instance, machine-learning-driven atomistic simulations [38] or big data analysis [39] have been applied to model battery materials. Herein, the advantage of machine learning in processing and analyzing large data sets or complex problems with nonlinear relationships can be exploited to assist with our data analysis in LiCoO$_2$. We therefore, utilize a typical machine learning method, namely, back propagation neural network (BPNN) to analyze the lithium-ion migration energy barriers in strained LiCoO$_2$.

## III. RESULT AND DISCUSSION

LiCoO$_2$ belongs to the α-NaFeO$_2$ type layered structure with the space group of R$\bar{3}$m [40-45]. In the unit cell of crystal structure, Li, Co and O atoms occupy the 3a, 3b and, 6c sites, respectively. Therefore, the Li layer and CoO$_2$ layer are alternately arranged along the *c*-axis direction, forming an ordered layered structure. Using aforementioned BMH potential, the optimized lattice constants of LiCoO$_2$ are found to be $a_0 = b_0 = 2.84$ Å and $c_0 = 13.92$ Å, which are consistent with experimental observations [43,46].

Lithium ions migrates to adjacent octahedral vacancies through one of two mechanisms, namely, oxygen dumbbell hop (ODH), and tetrahedral site hop (TSH), depending on lithium vacancy arrangement around the jumping ions [47,48]. The migration paths of these two mechanisms are shown in Fig. 1. The first transition mechanism occurs when lithium ions move along the linear (shortest) path connecting the jumping start point and the adjacent vacancy. The trail shown by the arrow in Fig.1 (a and b) passes through an oxygen ion dumbbell. This migration process is so-called oxygen dumbbell hop (ODH). When the local density of Li ion vacancy is low, Li ion migration by the ODH mechanism is dominated. However, with local density of Li ion vacancy increases, for example, as shown in Fig. 1(c and d), two adjacent vacancies of Li ion are created, then the lithium moves along a curved path, which passes through the tetrahedral position. Such migration path is referred as tetrahedral site hop (TSH). Similar to ODH and TSH, one can also model more possible migration paths by considering the cases with denser Li ion vacancies locally. It is worth pointing out that, here we emphasize on local vacancy density, to distinguish from the overall composition or overall vacancy. For example, for $Li_xCoO_2$, the overall composition of Li ions ($x$) can be very close to 1, however it is possible to have one or two vacancies locally in the region 1 of inner sphere in our defect calculations (as aforementioned in the method session). The ODH mechanism has been investigated by lattice dynamics simulation [30], first principles calculation [19,47,48] and Monte-Carlo simulations [47]. Nevertheless, the strain-dependent ion migration is much less studied. Although the Li ion migrations with in-plane and $c$-axis strains have been reported recently [19], more strain conditions such as pressure or combination of in-plane and $c$-axis strains are still lack of study. The TSH mechanism has been simulated by Monte-Carlo method [47]. However, there is still no yet report on TSH migration as a function of strains.

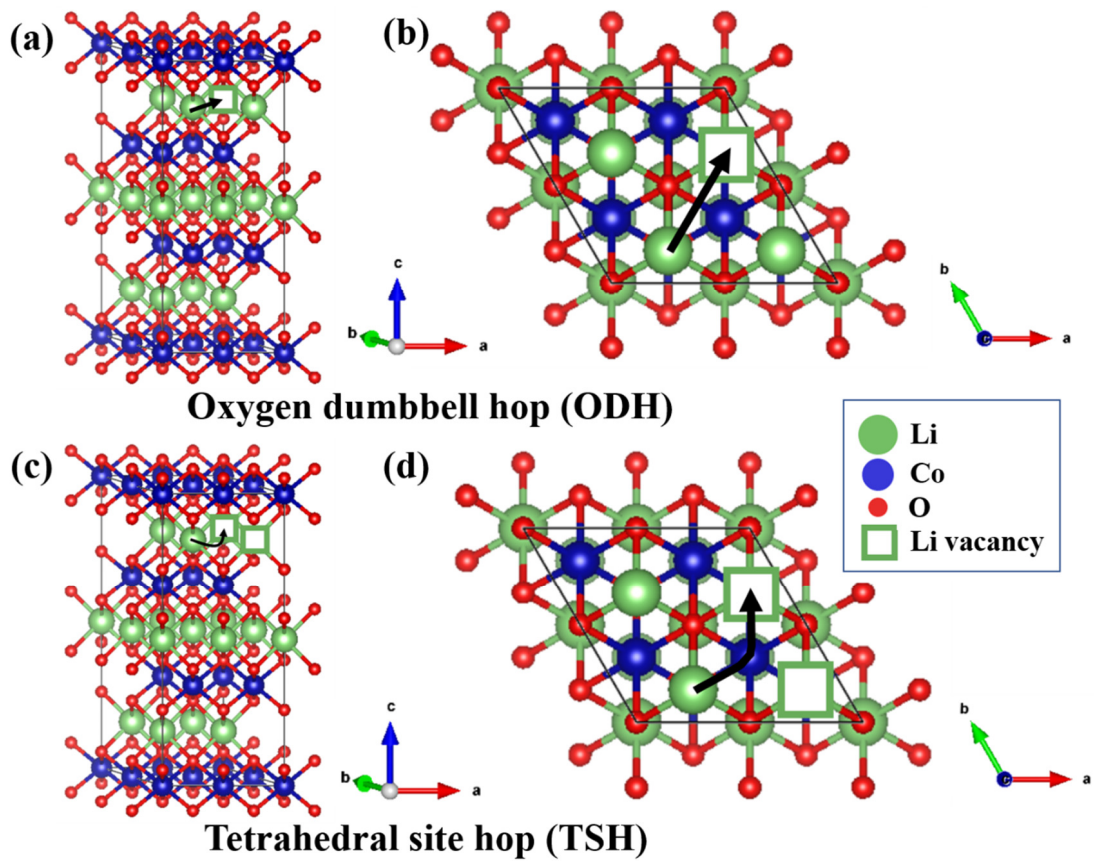

**Fig.** 1. Side and top views of the atomic strucutres of the two Li ion migration paths in $LiCoO_2$. (a,b) Oxygen dumbbell hop (ODH), (c,d) Tetrahedral Site Hop (TSH). The large (green), medium (blue), and small (red) spheres are Li, Co, and O atoms, respectively. The empty squares are lithium vacancies.

Firstly, we studied the migration energy barrier of three ions in $LiCoO_2$ without external strain. The migration energy barrier of Co is 2.89 eV, and that of O is 1.66 eV, respectively. The Li-ion migration energy barrier for the ODH mechanism is 0.49 eV, well reproducing the reported value of 0.48 eV by lattice dynamics simulation using the same BMH potential model [30], and it is also within the value of range (0.39 eV [19], 0.80 eV [47]) predicted by first principles calculations. The Li-ion migration energy barrier for the TSH mechanism is about 0.19 eV in our work, which is in good agreement with value of 0.23 eV (for $Li_xCoO_2$ with composition close to 1) predicted by first principles calculation [47]. It is interesting to find that the migration energy barrier for the TSH mechanism is much lower than that for the ODH mechanism. This finding has also been verified by Monte Carlo simulations [47].

Then we can use the calculated Li-ion migration energy barrier to estimate the diffusion coefficient. In the transition mechanism, the rate transition Γ can be calculated based on the transition state theory [49]. According to the lattice gas model [50], the diffusion coefficient can be calculated by

$$D = v_0 \exp\left(-\frac{E_b}{k_B T}\right) \times l^2 \qquad (2)$$

where $v_0$ is the vibration frequency of the migrating Li ions in the lattice, $E_b$ is the migration energy barrier, $k_B$ and $T$ are Boltzmann's constant and absolute temperature, respectively. The last quantity $l$ is the jumping distance, which is approximately equal to the lattice constant (in this study, we choose $l = 2.84$ Å, the lattice constant of $LiCoO_2$). The value of about 10 THz, as estimated from phonon frequency [19,51], is used for approximating vibration frequency $v_0$. With these parameters, the diffusion coefficient under zero strain at 300 K is estimated to be $4.7 \times 10^{-11}$ cm$^2$s$^{-1}$ for the ODH mechanism, which is in good agreement with the experimental observations [52]. The diffusion coefficient under zero strain at 300 K is estimated to be $5.19 \times 10^{-6}$ cm$^2$s$^{-1}$ for the TSH mechanism. The diffusion coefficient of TSH is much larger than that of ODH.

It can be seen that the lowering of migration energy increases diffusion coefficient significantly. Therefore, we then focus on how to lower the Li-ion migration energy barrier by considering the influence of strain on $LiCoO_2$ lattice. In order to investigate the effects of strain on ion transport, we systemically study the migration energy for any combination of strains $\Delta a/a$ ($\Delta b/b = \Delta a/a$) and $\Delta c/c$ in the range of (−3% ~ 3%), as shown in Fig. 2. Such map covers several typical strain conditions such as ideal cases of constant $c/a$ ratio ($c/a = c_0/a_0$), constant $a$ ($a = a_0$), constant $c$ ($c = c_0$), and constant volume V (V = $V_0$), as well as three typical realizable strain conditions, including $c$-axis uniaxial strain (1D strain), in-plane biaxial strain (2D strain), and hydrostatic pressure (3D strain).

We found that by increasing $\Delta a/a$ and $\Delta c/c$, the energy barrier of lithium-ion migration for the ODH mechanism can be reduced. However, for TSH, the migration energy barrier can be lowered by decreasing $\Delta a/a$ and increasing $\Delta c/c$. It is interesting to find that the better location in the strain map with lower migration energy barrier is in the region of positive $\Delta a/a$ and positive $\Delta c/c$ for ODH (up-right corner in Fig.2 (a)), but in the region of negative $\Delta a/a$ and positive $\Delta c/c$ for TSH (up-left corner in Fig.2 (d)). The common feature for the reduction of migration energy for both ODH and TSH cases is the increase of $\Delta c/c$, namely, enlargement of $c$ lattice. The reason for the significant decrease of migration energy barrier is that a larger interlayer spacing is favorable for Li migration, considering the 2D transport nature of Li ions in layer-structure LiCoO$_2$. However, the $E_b$ responses to $\Delta a/a$ are rather different between ODH and TSH. Since the main difference in the structural feature between ODH and TSH is the density of local vacancies of Li ion, our results suggest that the migration energy barrier of Li ions is very sensitive to local vacancy density. During charging process, more and more Li ions are removed from LiCoO$_2$, thus the local vacancy density is likely to increase. The switching from ODH to TSH with increasing local Li vacancy density is then expected. Moreover, from realistic point of view, the strain condition with negative $\Delta a/a$ and positive $\Delta c/c$ (up-left corner in Fig.2 (d)) should be easier to be achieved comparing to the case with both positive $\Delta a/a$ and $\Delta c/c$ (up-right corner in Fig.2 (a)), considering the positive Poisson ratio of LiCoO$_2$. Therefore, from the screening of $E_b$-strain map, the possible to-be-achieved location of low migration energy barrier of strained LiCoO$_2$ seems very promising. Of course, there will be also some possible extreme cases such as volume expansion during charging-discharging process, which might be likely to have the strain condition with both positive $\Delta a/a$ and $\Delta c/c$.

To explore as many strain conditions as possible, and also to make the results of migration energy barrier more useful, we adopt an empirical formula to fit the $E_b$-strain data. The obtained empirical formula is expected to be conveniently applied in experimental cases with available measured strain conditions or strain fields. Based on the observation of the trend of $E_b$-strain relationship, as presented in Fig. 2(a), and for keeping simplification, we used a 2D coupled quadratic function to fit the data, that is,

$$E_b(x,y) = E_{b0} + \alpha_1 x + \alpha_2 x^2 + \beta_1 y + \beta_2 y^2 + \lambda xy \qquad (3)$$

where $E_{b0}$ is the migration energy barrier without strain, $x = \Delta a/a$, $y = \Delta c/c$, $\alpha_i$, $\beta_i$, and $\lambda$ are fitting parameters. For the ODH mechanism, we obtained the following fitting parameters: $E_{b0}$ = 0.49 eV, $\alpha_1$ = -3.99223, $\alpha_2$ = -11.61337, $\beta_1$ = -6.05768, $\beta_2$ = 17.42128, and $\lambda$ = 12.16358. The fitting results are presented in Fig. 2(b). For TSH, the fitting parameters are given as follows: $E_{b0}$ = 0.19 eV, $\alpha_1$ = 0.42798, $\alpha_2$ = 3.22017, $\beta_1$ = -6.60335, $\beta_2$ = 74.98096, and $\lambda$ = 39.16995, which lead to a good fitting as shown in Fig. 2(e).

To further analyze the data, we utilize BPNN, one of the machine learning methods on data fitting of migration energy barrier. The fitting results are shown in Fig. 2(c) for ODH and Fig. 2(f) for TSH, respectively, which exhibit excellent fitting to LD simulations. We expect the usage of machine learning method may be even more useful for optimization with more variances, or more complex cathode materials such as LiCo$_x$Mn$_y$Ni$_z$O$_2$ with a huge number of local structural configurations.

The useful feature of the analytic formula or machine learning model is that once the strain conditions ($\Delta a/a$ and $\Delta c/c$) of LiCoO$_2$ are known (either from theoretical calculations such as first principles calculations, or from experimental measurement such as x-ray diffraction or electron diffraction), the migration energy barrier $E_b(x,y)$ can be directly and quickly estimated, with no need of further expensive LD simulation or first principles calculations.

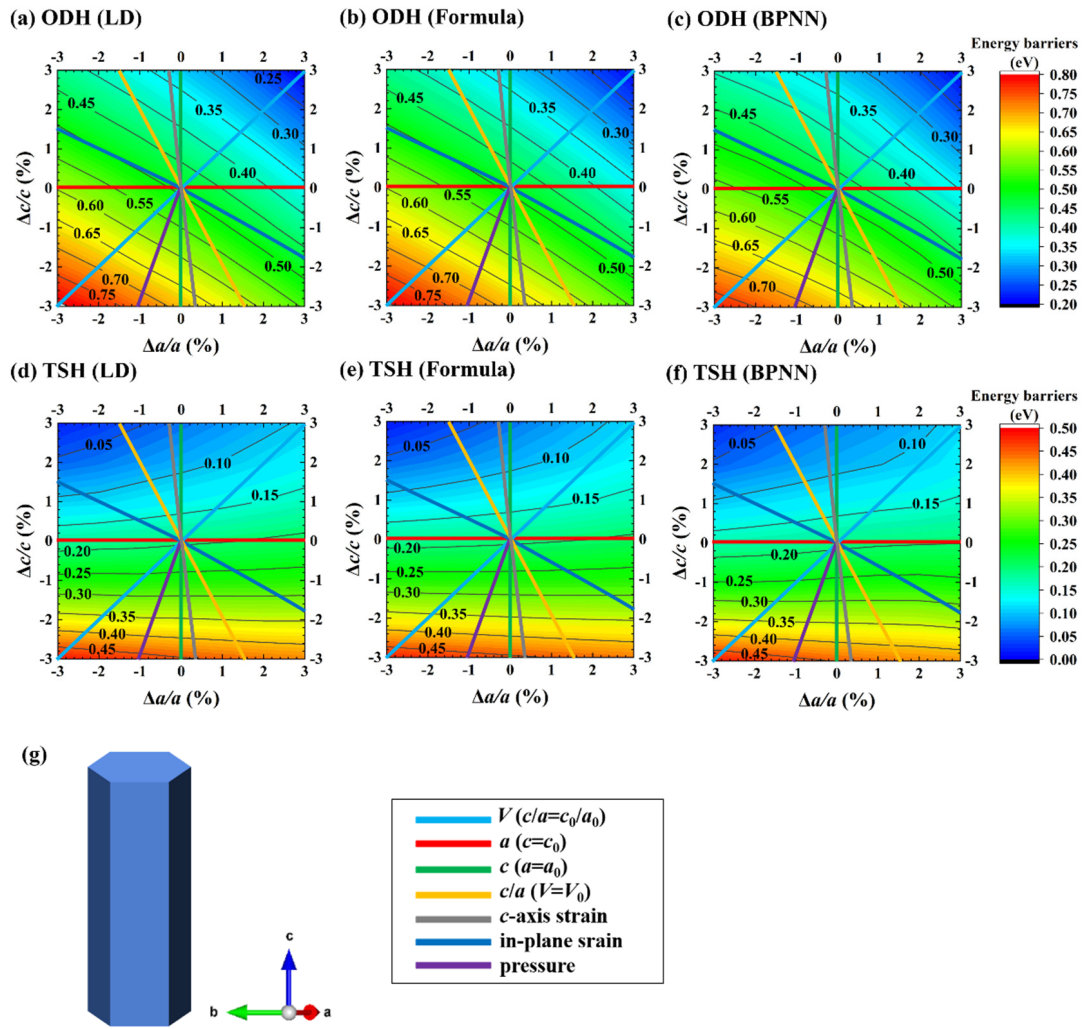

FIG. 2. 2D contour plot of the Li-ion migration energy barrier of LiCoO$_2$ under external strain. The Li-ion migration energy barriers $E_b$ obtained from (a) lattice dynamics calculation (labeled as LD), (b) by fitting data using empirical formula, Eq.(3) (labeled as Formula), and (c) by fitting data using machine learning method, namely, BP neural network (labeled as BPNN) for the ODH mechanism. (d), (e) and (f) are for TSH case and the corresponding labels are similar to (a), (b) and (c), respectively. (g) Indication of directions of strain axis in LiCoO$_2$. The colored bold lines indicate different strain conditions in LiCoO$_2$. (from top to bottom, blue: volume variation $V$ with fixed $c/a$; red: variation of lattice constant $a$ with fixed $c$; green: variation of constant lattice $c$ with fixed $a$ and $b$; yellow: variation of $c/a$ with fixed volume; gray: lattice variation for $c$-axis strain; dark blue: lattice variation for in-plane strain; purple: lattice variation for pressure up to 10 GPa).

The foregoing results and discussions are for general strain conditions with any arbitrary combination of strains $\Delta a/a$ ($\Delta b/b = \Delta a/a$) and $\Delta c/c$. It forms the basic and complete database for further detailed data analysis. We then focus on three typical strain conditions that are often applied in materials science, *i.e.*, $c$-axis uniaxial strain (1D strain), in-plane biaxial strain (2D strain), and hydrostatic pressure (3D strain). We start from the hydrostatic pressure case, which contain only one variance, namely, the volume of LiCoO$_2$. The migration energy barriers of ODH and TSH of Li ions in LiCoO$_2$ as a function of pressure are presented in Fig. 3. Similar results are observed for LD simulation, analytic formula, and BPNN, indicating the good fitting of analytic formula (Eq. 3) and machine learning model.

When pressure ($P$) is applied, both the lattice constants $a$ ($b$) and $c$ are compressed (red line in Fig. 2), and $E_b$ increases with the increase of the pressure (decrease of volume), as shown in Fig 3. It is interesting to see that the $E_b$-$P$ relationship is almost linear, with evidence from good fit of $E_b$ as a function of $P$,

$$E_b(P) = E_{b0} + \gamma P \qquad (4)$$

with linear coefficient $\gamma = 0.02412$ for ODH and $\gamma = 0.02653$ for TSH, respectively. One can also fit the curve with a 2$^{nd}$ order polynomial function, namely, $E_b(P) = E_{b0} + \gamma_1 P + \gamma_2 P^2$, with $\gamma_1 = 0.02321$, and $\gamma_2 = 1.15895 \times 10^{-4}$ for ODH and $\gamma_1 = 0.02196$, and $\gamma_2 = 5.81617 \times 10^{-4}$ for TSH, respectively. Again, the small values of $\gamma_2$ confirm the linearity of $E_b$-$P$ relationship. The other interesting feature can be observed from Fig. 3 is that the slopes of $E_b$ as a function of $P$ are very close to each other for ODH and TSH. The overall contour features of $E_b$ vs $\Delta a/a$ and $\Delta c/c$, as shown in Fig. 2, are quite different between ODH and TSH, especially, the top panel. However, the bottom-left section (regions with both negative $\Delta a/a$ and $\Delta c/c$) of Fig. 2(a) and Fig. 2(d) are sharing very similar tomographic feature, resulting in the similar $E_b$-$P$ trend.

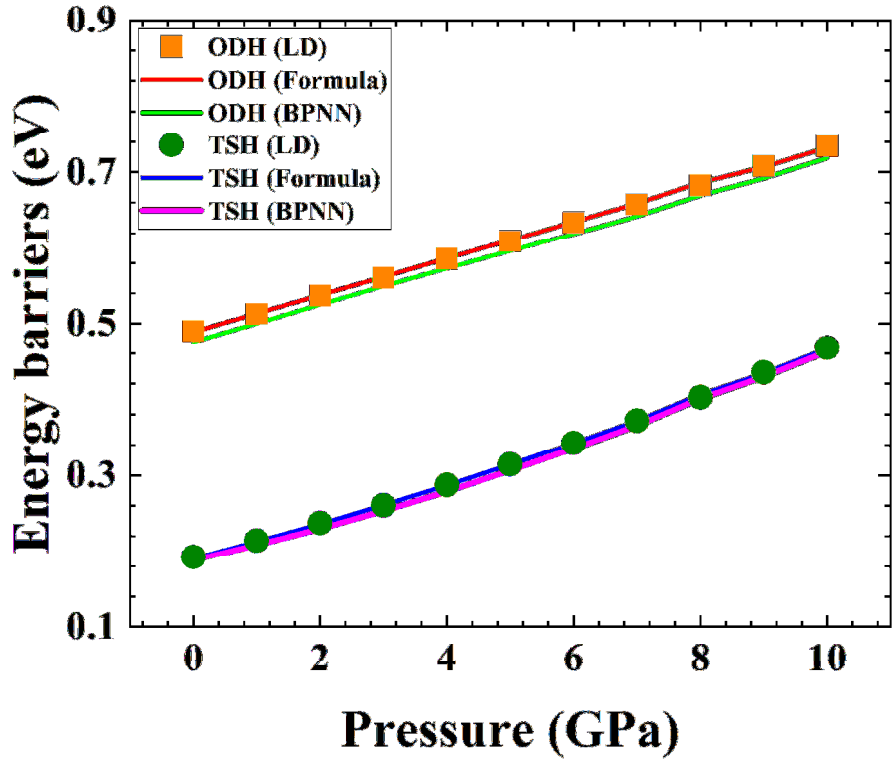

Fig.3. Li-ion migration energy barrier of ODH and TSH in LiCoO$_2$ as a function of pressure. Results obtained from LD simulation, analytic formula, and BPNN are also listed for comparison.

From curves in Fig. 3 and above fitting results, we concluded that applying pressure always give rise to $E_b$. This is because the lattice shrinks in three dimensions under pressure, which results in smaller spaces for Li ion to move, and shorter interatomic spacing will enhance the repulsive interaction between Li ions with $CoO_2$ planes and adjacent Li ions, thus leads to higher migration energy barrier.

Finally, we examined two typical strain conditions, namely, *c*-axis strain and in-plane strain, which can be realized in the thin film of $LiCoO_2$ battery. The uniaxial *c*-axis strain is to compress or stretch lattice *c* at certain values and other lattice parameters *a* and *b* are allowed to relax. While for in-plane strain, the lattice parameters *a* and *b* are changed at the same time, and lattice parameter *c* is relaxed accordingly. In this way, the uniaxial *c*-axis strain is nearly 1D strain, and in-plane strain can be regarded as 2D strain, compared with pressure (3D strain). As shown in Fig. 2 (gray line) and Fig. 4 (a), whether it is for the ODH or TSH mechanism, under strain along the *c*-axis direction, the Li-ion migration energy barrier increases with compressive strain and decreases with tensile strain. That is, when the *c*-axis strain increases from -3% to 3%, the Li-ion migration energy barrier continues decreasing monotonically. Moreover, the Li-ion migration energy barrier of TSH is much lower and decreases much faster than that of ODH when the lattice parameter *c* is increasing. This makes the tensile *c*-axis strain a very promising tunable way to reduce energy barrier. The trend of variation of Li-ion migration energy barrier as a function of *c*-axis strain is similar with previous first principles calculations [19], but the findings of strain-dependent TSH are first reported in this work, according to the best of our knowledge.

Interestingly, the response of energy barrier to in-plane strain is quite different from *c*-axis strain, both for ODH and TSH. As shown in Fig. 2 (dark blue line) and Fig. 4 (b), in the *a-b* basal plane (in-plane) under biaxial strain, the Li ion migration energy barrier for the ODH increases upon compressive strain while decreases upon tensile strain conversely, but only have small variations (< 0.04eV). However, for the TSH, the Li-ion migration energy barrier increases upon tensile strain while decreases upon compressive strain conversely. When the tensile strain is applied in the *a-b* basal plane, it can be also observed that the Li-ion migration energy barrier difference between the ODH and TSH keeps shrinking with the increases of tensile strain.

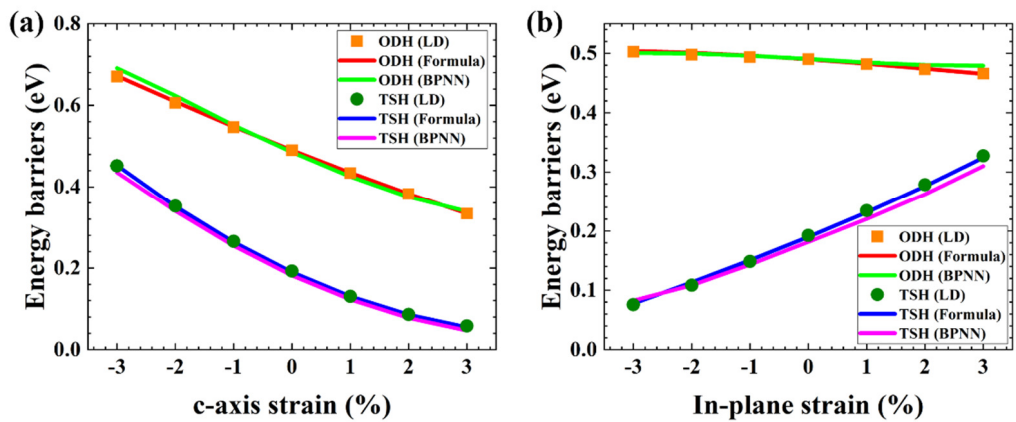

Fig. 4. The Li-ion migration energy barrier for ODH and TSH in $LiCoO_2$ as a function of strain. (a) the uniaxial strain along *c*-axis; (b) biaxial strain in the *a-b* basal plane.

Our striking results reveal the possibility of applying strain to reduce Li-ion migration energy barrier further, for example, tensile uniaxial *c*-axis strain for ODH and TSH cases or compressive in-plane strain for TSH case (as shown in Fig. 4). Generally speaking, the presence of nanoparticles may produce a local strain field (see for example [53]). As long as the local strain field can be determined, the change in lattice constants can be measured experimentally, (e.g., by determination of x-ray scattering factor using synchrotron x-ray diffraction or electron scattering factor using electron diffraction [54-57]); one can use the empirical formula or machine learning method, as presented in this work, to estimate Li-ion migration energy barrier. On the other hand, to generate desired strain with lower Li-ion migration energy barrier, one may also consider other tuning ways such as control of grain boundary or doping.

## IV. CONCLUSION

In summary, we have performed LD simulations, empirical formula analysis, and machine learning method to study Li-ion migration energy barriers in $LiCoO_2$ under different strain conditions. We have identified two main migration mechanisms, namely, ODH and TSH in $LiCoO_2$ with different local Li vacancy density. Large database containing Li-ion migration energy barriers for general strain conditions with any arbitrary combination of strains $\Delta a/a$ ($\Delta b/b = \Delta a/a$) and $\Delta c/c$ are presented and three typically realizable strain conditions including applying *c*-axis uniaxial strain (1D strain), in-plane biaxial strain (2D strain), and hydrostatic pressure (3D strain) are analyzed in great details. We deduced the general trend of the strain-dependent energy barrier of lithium-ion migration in $LiCoO_2$ in an analytic empirical form. According to this formula, one can estimate the Li ion energy barrier with available lattice parameters or pressure. We found that it is possible to further reduce Li-ion migration energy barrier with applied tensile uniaxial *c*-axis strain for ODH and TSH cases or compressive in-plane strain for TSH case. The strain engineering method proposed here aims to reduce the Li-ion migration energy barrier and may be applied to various strain engineering aspects of other functional materials.

## Acknowledgement

This work is supported by XMUM Research Fund XMUMRF/2019-C3/IORI/0001.